\documentclass[useAMS,usenatbib]{mnras}
\usepackage{times}
\usepackage{epsf}
\usepackage{hyperref}

\newcommand{\ellq}{l}
\newcommand{\dd}{\mathrm d}

\title[Time-varying solar surface term]{Parametrizing the time-variation of the `surface term' of stellar p-mode
frequencies: application to helioseismic data
}
\author[R. Howe et al.]{R. Howe$^{1}$\thanks{E-mail:
rhowe@nso.edu (RH)}
S. Basu$^2$, 
G.~R.~Davies$^{1,3}$,
W.~H.~Ball$^{4,5}$, 
W.~J.~Chaplin$^{1,3}$, Y.~Elsworth$^{1,3}$, R.~Komm$^6$\\ 
$^{1}$School of Physics and Astronomy, University of Birmingham, Edgbaston, Birmingham B15 2TT, United Kingdom\\
$^2$Department of Astronomy, Yale University, PO Box 208101, New Haven, CT
06520-8101, USA
\\
$^{3}$Stellar Astrophysics Centre (SAC), Department of Physics and As
tronomy, Aarhus University, Ny
Munkegade 120, DK-8000 Aarhus C, Denmark\\
$^{4}$Institut f{\"u}r Astrophysik, Georg-August-Universität G{\"o}ttingen, Friedrich-Hund-Platz 1, 37077 G{\"o}ttingen, Germany\\
$^5$Max-Planck-Institut f{\"u}r Sonnensystemforschung, Justus-von-Liebig-Weg 3, 37077 G{\"o}ttingen, Germany\\
$^6$National Solar Observatory, 950 N. Cherry Avenue, Tucson, AZ 85719, USA}

\begin{document}



\maketitle

\label{firstpage}

\begin{abstract}
The solar-cyle variation of acoustic mode frequencies has a frequency dependence related to the inverse mode inertia. The discrepancy between model predictions and measured oscillation frequencies for solar and solar-type stellar acoustic modes includes a significant frequency-dependent term known as the surface term that is also related to the inverse mode inertia. We  parametrize both the surface term and the frequency variations for low-degree solar data from Birmingham Solar-Oscillations Network (BiSON) and medium-degree data from the Global Oscillations Network Group (GONG) 
using 
the mode inertia  together with cubic and inverse frequency terms. 
We find that for the central frequency of rotationally split multiplets the cubic term dominates both the average surface term and the temporal variation, but for the medium-degree case the inverse term improves the fit to the temporal variation. We also examine the variation of the even-order splitting coefficients for the medium-degree data and find that, as for the central frequency, the latitude-dependent frequency variation, which reflects the changing latitudinal distribution of magnetic activity over the solar cycle, can be described by the combination of a cubic and an inverse function of frequency scaled by inverse mode inertia. The results suggest that this simple parametrization could be used to assess the activity-related frequency variation in solar-like asteroseismic targets.

\end{abstract}

\begin{keywords}
Sun: helioseismology -- Sun: oscillations.
\end{keywords}

\section{INTRODUCTION}

It is well established that the solar oscillation mode frequencies show a small variation with activity level; see, for example, \citet{1985Natur.318..449W,1989MNRAS.238..481C,1990Natur.345..322E,1994ApJ...434..801E,2004A&A...413.1135S,2007ApJ...659.1749C} for low-degree modes and \citet{1990Natur.345..779L,1999ApJ...524.1084H} for medium-degree modes. Similar variations may also be seen in asteroseismic targets that have a solar-like oscillation spectrum and 
an activity cycle \citep[see, for example][]{2007MNRAS.377...17C,2010Sci...329.1032G}.
This variation has a frequency dependence below the acoustic cut-off frequency that implies it is dominated by near-surface effects \citep{1990Natur.345..779L,1994ApJ...434..801E}. \cite{1990Natur.345..779L} found that the 
the shift in medium-degree mode frequencies varied as the 
cube of the frequency divided by the mode inertia, and \cite{1991ApJ...370..752G} provided some theoretical justification for this in terms of a photospheric perturbation. \citet{1990LNP...367..283G}
proposed parametrizing the frequency dependence using a cubic and an inverse term in frequency, both scaled by the inverse mode inertia.  The cubic term is potentially related to the presence of something, perhaps magnetic flux tubes, that affects sound speed but not density, while the inverse term would reflect something like an error in the pressure scale height near the surface.

This two-term parametrization was recently applied by \citet{2014A&A...568A.123B} to describe the frequency dependence of the so-called `surface term'
in low-degree solar and asteroseismic data. This term is the 
frequency-dependent correction that is applied in solar and stellar seismology to account for differences between model and observed frequencies that arise from deficiencies in modelling the near-surface layers.  They concluded that for asteroseismic uses the cubic term alone was adequate, at least for a reasonably Sun-like star. On the other hand, \citet{2015ApJ...808..123S} found that for the model to apply across the H--R diagram, the inverse term was also needed.


In this work we return to the idea of \citet{1990LNP...367..283G} and apply the two-term model to the solar-cycle frequency shifts in low- and medium-degree solar p modes. This will both test the hypothesis that the variation will be dominated by the cubic term (reflecting a change in something that behaves like a fibril magnetic field) and provide a convenient way to summarize the variation over a wide range of frequency and degree.

\section{DATA AND ANALYSIS}
\subsection{Helioseismic Data}
The Birmingham Solar-Oscillations Network (BiSON) has been collecting Sun-as-a-star helioseismic data since the late 1970s, and since 1993 it has operated as a six-site network with an average duty cycle of about 80 per cent \citep{2016SoPh..291....1H}. 
The BiSON Doppler velocity residuals were prepared as described in \cite{2014MNRAS.441.3009D} and divided into 23 non-overlapping time series each covering one year, commencing with 1993 and ending with 2015. We reused the fitted parameters generated for the analysis of \citet{2015MNRAS.tempH}, using a maximum likelihood estimation algorithm based on that of \citet{1999MNRAS.308..424C}, and we extended the dataset by including the 2015 observations.

The Global Oscillations Network Group (GONG) began resolved-Sun helioseismic observations from a six-site network in 1995 \citep{1996Sci...272.1284H}. The GONG frequencies used here were those generated by the GONG project pipeline \citep{1996Sci...272.1292H}\footnote{\url{http://gong.nso.edu}}, based on  68 non-overlapping time series of 108 days beginning in May 1995. The GONG analysis provides a frequency for each mode of azimuthal order $m$ within a multiplet of radial order $n$ and degree $l$, up to $l=150$;  these were then fitted to a polynomial expansion in $m$ to obtain the central frequency and the splitting coefficients.

We note that the BiSON data were fitted with an asymmetric peak profile model while the medium-degree data were fitted with a symmetric one. As was discussed by  \citet{2015MNRAS.tempH}, this makes little or no difference to the determination of the temporal variation of the frequencies.

\subsection{Activity Indices}

As in \citet{2015MNRAS.tempH}, for a global activity index we use the daily 10.7\,cm Radio Flux (RF) index averaged over the observing period corresponding to each frequency data set.

As there is no latitudinal information in the RF index, we use a magnetic 
proxy for the two-dimensional analysis. In this work we use the National Solar Observatory (NSO) Kitt Peak synoptic charts\footnote{\url{http://nsokp.nso.edu/dataarch.html}} and the NSO {\it Synoptic Optical Long-term Investigations of the Sun} (SOLIS)  synoptic maps\footnote{\url{http://solis.nso.edu/solis_data.html}}. We convert the SOLIS synoptic maps of magnetic flux \citep{2011SPIE.8148E..09B,2013SoPh..282...91P} to absolute values (in gauss, G) and bin them.
We have filled eleven gaps with GONG magnetograms, to which we applied scaling factors.  The scaling factors were derived from a regression analysis
between SOLIS and GONG magnetograms and are 1.63 for Carrington rotation (CR)~2059 and 2063, 1.5 for 
CR~2090 and 2091, and 1.3 for CR~2152\,--\,2155, 2163, 2166, and 2167.  We then form a global magnetic index by taking the mean of the  absolute field strength across all of the latitude bins for each rotation.

\subsection{Analysis}
\subsubsection{Construction of the surface term}
The surface term was approximated by taking the difference $\delta\nu_{nl}$ between the observed frequencies and the much-used `Model S' of  \citet{1996Sci...272.1286C}. This was then scaled by
the `mode inertia', $E_{nl}$, also taken from Model S and normalized to give a value of unity at 3\,mHz for $l=0$.

\subsubsection{Parametrization}

We then fit the scaled frequency differences, $E_{nl}\delta\nu$, to a function 
\begin{equation}
{F(\nu)}={a_{\mathrm{inv}}({\nu\over{\nu_{\mathrm{ac}}}})^{-1}+a_{\mathrm{cubic}}({\nu\over{\nu_{\mathrm{ac}}}})^3},
\label{eq:eq1}
\end{equation}
where $\nu_{\mathrm{ac}}$ is the acoustic cut-off frequency, which is here taken to be
5\,mHz \citep{2011ApJ...743...99J} and is used to ensure that both coefficients have the same dimensions; this means that both coefficients have the same units of frequency as the LHS of Equation~(\ref{eq:eq1}). $a_{\mathrm{inv}}$ and $a_{\mathrm{cubic}}$ are the coefficients to be fitted. Note that this expression does not include a constant term. For comparison, we also fit expressions with only the $a_{\mathrm{cubic}}$ or $a_{\mathrm{inv}}$ terms.
The full calculation for one-term and two-term linear least-squares fits is given in Appendix~\ref{sec:app}.

 We note that the value chosen for $\nu_{\mathrm{ac}}$ will change the numerical values of the coefficients in the two-term fit but has no impact on the goodness of fit or on the model prediction for the frequency shift.

\section{RESULTS}

\subsection{Central frequency}
\label{sec:cent}

\subsubsection{Stationary part and sensitivity}

We start by assuming that the activity-related change in $E_{nl}\delta\nu$ is a linear function of the global activity index.  As we shall see later, this is only an approximation, but it is a reasonable one for the purpose of a first look at the data. We can then perform a linear regression for the scaled frequency of each mode, using the activity proxy as the independent variable, to give a `zero-activity' value (the intercept of the fit) and a `sensitivity' value (the slope of the fit, which for convenience we sometimes refer to as $\dd\nu/\dd RF$). 

\begin{figure*}
\epsfxsize=0.8\linewidth\epsfbox{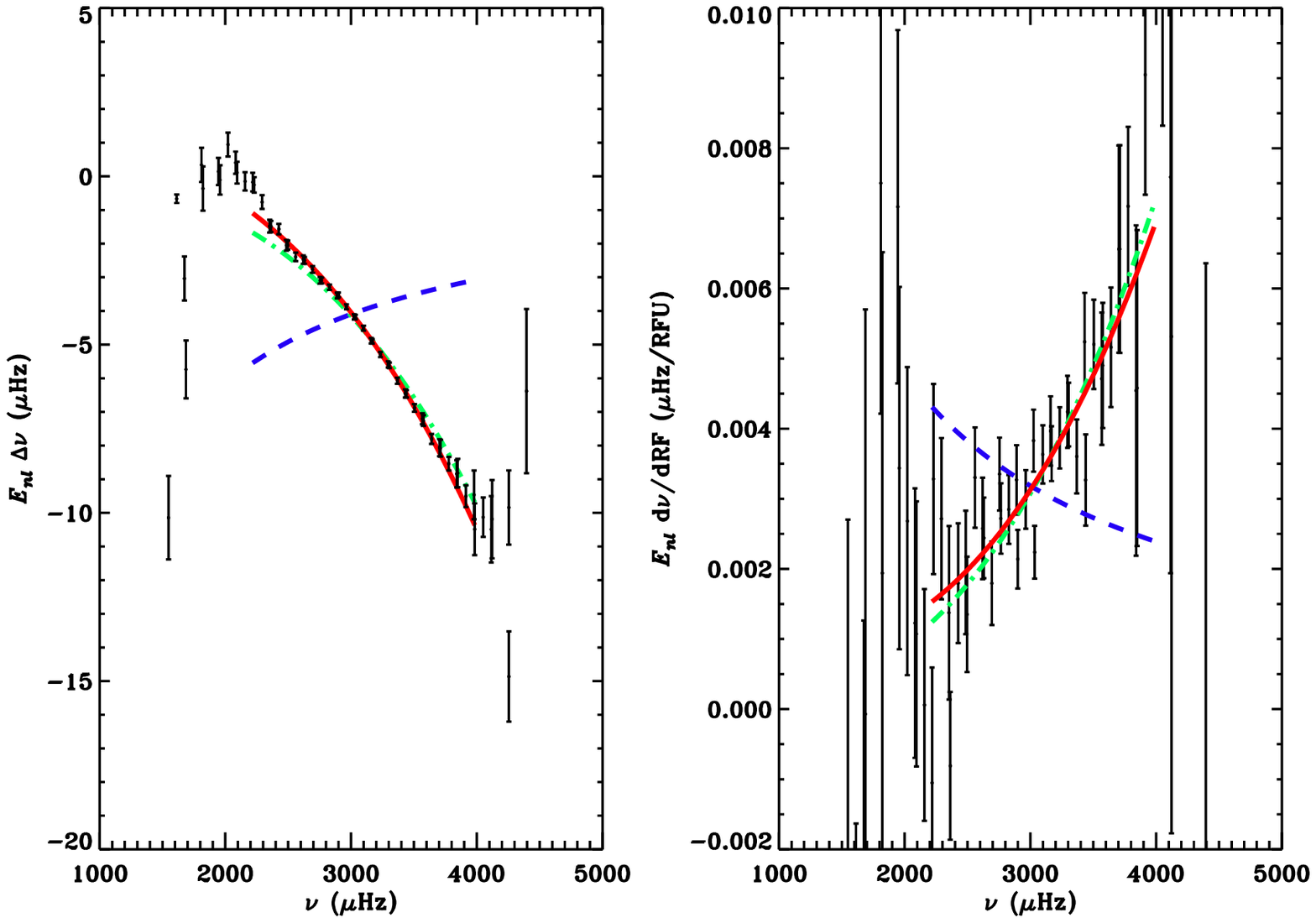}
\epsfxsize=0.8\linewidth\epsfbox{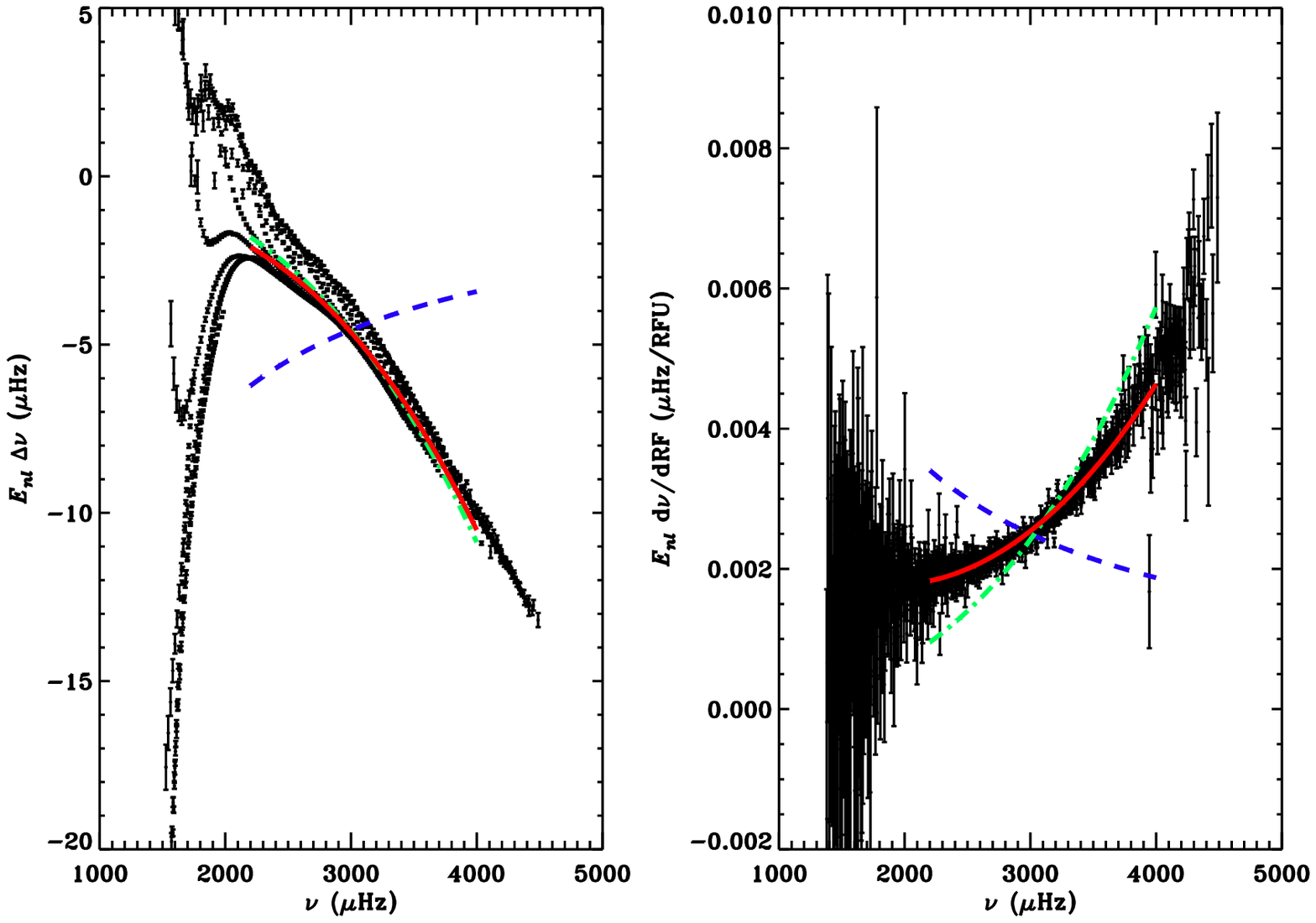}
\caption{Intercept (left) and slope (right) of a linear fit of the scaled differences between the BiSON (top) and GONG (bottom) frequencies and Model S to the RF index. The solid curve shows the best two-term fit, the dot-dashed curve the best fit with the cubic term only, and the dashed curve the best fit with inverse term only. Fits were carried out over the frequency range from $2200$ to $4000\,\mu{\mathrm{Hz}}$. For clarity, we show 5-$\sigma$ error bars on the intercept plots and 1-$\sigma$ on the gradient.}
\label{fig:fig1}
\end{figure*}

The zero-activity and sensitivity values for $E_{nl}\delta\nu$ are plotted in Figure~\ref{fig:fig1} for BiSON and 
 GONG. In the GONG case, modes of different radial order $n$ fall along slightly different curves in the zero-activity plot, reflecting the depth-dependent part of the
discrepancy from the model. To isolate the true surface term in the medium-degree case would require extra steps in the analysis; see, for example, \citet*{1989MNRAS.238..481C} for a method of preparing data for an asymptotic sound-speed inversion by fitting spline functions in frequency and in the depth-dependent term $\nu/(l+1/2)$. However, we see that the sensitivity shows little if any depth dependence, consistent with the understanding that the frequency shifts are confined to the near-surface layers, so as we are mostly interested in the temporal variation we choose to neglect the depth effects.

Overplotted on the data in Fig.~\ref{fig:fig1} 
are curves representing fits to the full two-term expression in Equation~(\ref{eq:eq1}) and to expressions using only the $a_{\mathrm{cubic}}$ or $a_{\mathrm{inv}}$ terms. In both cases, we see that the cubic term alone is much closer to the data than the inverse term alone, but that the combination of both terms gives a better fit than either alone. In quantitative terms, for the BiSON stationary term, the ratio between the $\chi^2$ values for the cubic-only fit compared to the two-term fit is    4.463, while for the inverse-only fit the ratio is  283.533. For the sensitivity term, the corresponding ratios are    0.991 and    2.968. The 0.1 per cent significance threshold is    2.738. For the GONG stationary term, the ratio between the $\chi^2$ values for the cubic-only fit compared to the two-term fit is    1.078, while for the inverse-only fit the ratio is   22.128. For the sensitivity term the corresponding ratios are   15.294 and   64.362; the 0.1 per cent significance threshold is    1.228. So we see that for BiSON both terms are significant in the stationary part, while only the cubic term is significant for the sensitivity. For GONG, on the other hand, the inverse term is not making a significant contribution to the stationary part, where the fit is poor because of the depth-dependent structure, but both terms are significant for the sensitivity. In all cases the inverse-only fit is bad.
The fitting range was restricted to $(2200 \leq \nu \leq 4000)\,\mu{\mathrm{Hz}}$. Below this range the stationary part of the frequency difference from the model diverges strongly from the simple description, particularly for the lower radial orders of the medium-degree data where the depth-dependent part becomes more important than the surface term; the upper limit to the fit was chosen to avoid outliers due to low signal-to-noise at higher frequencies.  The relatively poor fit to the stationary term at low frequencies reveals the limitations of this simple model;  however, we are concerned here primarily with the temporal variation, where the two-term expression is a reasonable description of the data albeit with some scatter at the low-frequency end; there is little or no order-dependent structure in the sensitivity plots.

 We note that there is not exact agreement between the two datasets even at low degree. See Section~\ref{sec:sectoral} for more discussion of this issue.

\begin{figure*}
\epsfxsize=0.4\linewidth\epsfbox{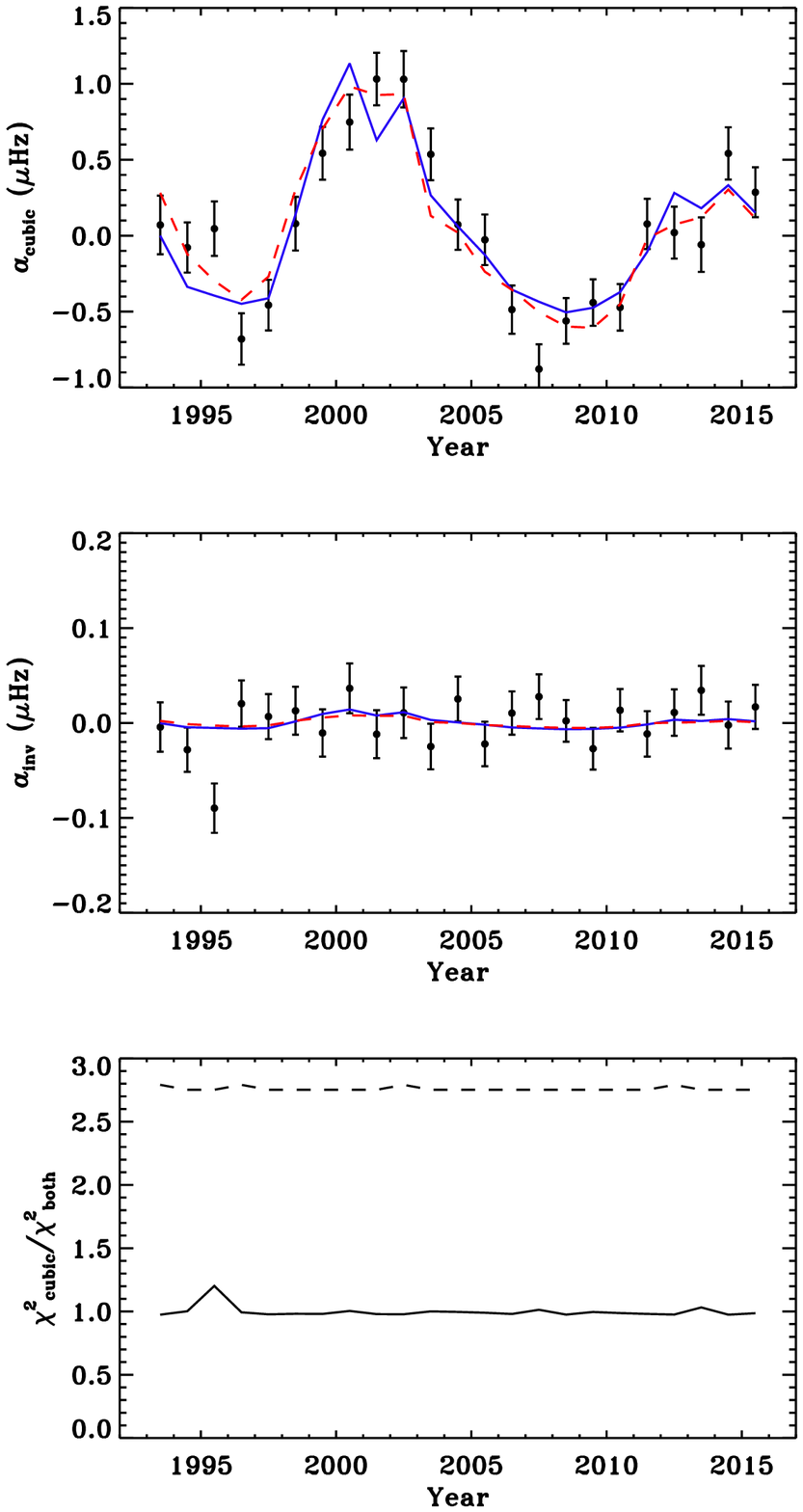}
\epsfxsize=0.4\linewidth\epsfbox{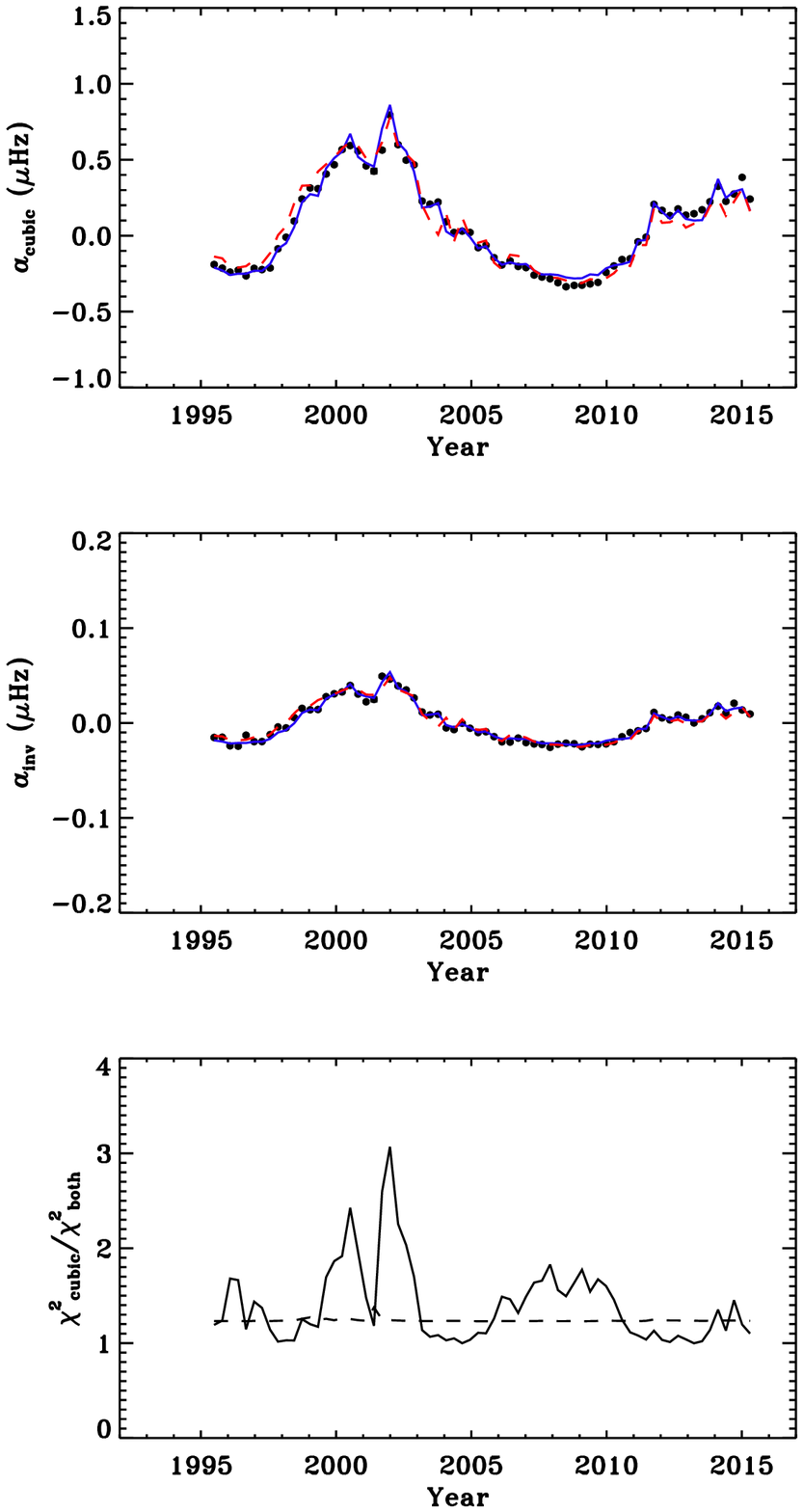}
\caption{Parametrization of scaled frequency shifts with respect to the mean frequency for each mode, for BiSON (left) and GONG (right). The  curves represent the best linear fits of the cubic and inverse terms  to the RF index (solid curve) and the global magnetic index (dashed curve). Top row: cubic term from two-term fit. Middle row: inverse term. Bottom row: $\chi^2$ ratio, $\chi^2_{\mathrm{cubic}}/\chi^2_{\mathrm{cubic+inv}}$, with the dashed line representing the 0.1 per cent significance threshold.} 
\label{fig:fig3}
\end{figure*}

\subsubsection{Temporal variation}

Next, to avoid the potential problems introduced by assuming a linear relationship between frequency and activity, we considered the difference from the mean weighted frequency over all the available time intervals for each mode in each time interval, the means being weighted by the inverse square of the uncertainty. We fit these differences with a two-term and a cubic-only fit. In Figure~\ref{fig:fig3} we show the temporal variation of the inverse and cubic terms for the two-term fit. To assess the significance of the additional inverse term we also show the ratio $\chi^2_{\mathrm{cubic}}/\chi^2_{\mathrm{cubic+inv}}$ and the 0.1 per cent signficance limit. While for the BiSON data this ratio falls well below the significance threshold, for the GONG data it rises above the threshold at some epochs, most notably during the
Cycle 23 maximum. These epochs appear to correspond to times when the frequencies are most different from the mean, or in other words those times when the 
signal-to-noise of the differences is high; this suggests that the medium-degree data should be fitted with the two-term expression throughout. While the $a_{\mathrm{inv}}$ coefficient 
is numerically about a factor of 15 smaller  than the $a_{\mathrm{cubic}}$ coefficient, we note that the relative contribution of the inverse term is larger than this; at 2.5 mHz ($\nu/\nu_{\mathrm{ac}}=0.5$) the two terms contribute about equally to the final value, while at 4.0 mHz ($\nu/\nu_{\mathrm{ac}}=0.8$) the contribution from the cubic term is larger by about a factor of 6. This implies that the inverse term plays a larger role in the fit
at the lower end of the frequency range, which corresponds to the modes with slightly deeper upper turning points.

Also shown are curves representing the best linear fit of the $a_{\mathrm{cubic}}$ terms to both the RF index and the global unsigned magnetic index. %
For BiSON, $a_{\mathrm{cubic}}$ from the two-term fits has correlation coefficients of 0.892 for 23 data pairs with the RF index and 0.919 with the Kitt Peak magnetic index,
while $a_{\mathrm{inv}}$ (which for the BiSON data appears to be largely noise) has (non-significant) correlation coefficients of 0.250 with RF and 0.173 with the magnetic index. For GONG, $a_{\mathrm{cubic}}$ from the two-term fits has correlation coefficients of 0.991 with the RF index and 0.978 with the Kitt Peak magnetic index,
while $a_{\mathrm{inv}}$ has correlation coefficients of 0.989 with RF and 0.973 with the magnetic index for the 68 independent datasets. 

\subsubsection{The sectoral mode effect}
\label{sec:sectoral}
\begin{figure*}
\epsfxsize=0.8\linewidth\epsfbox{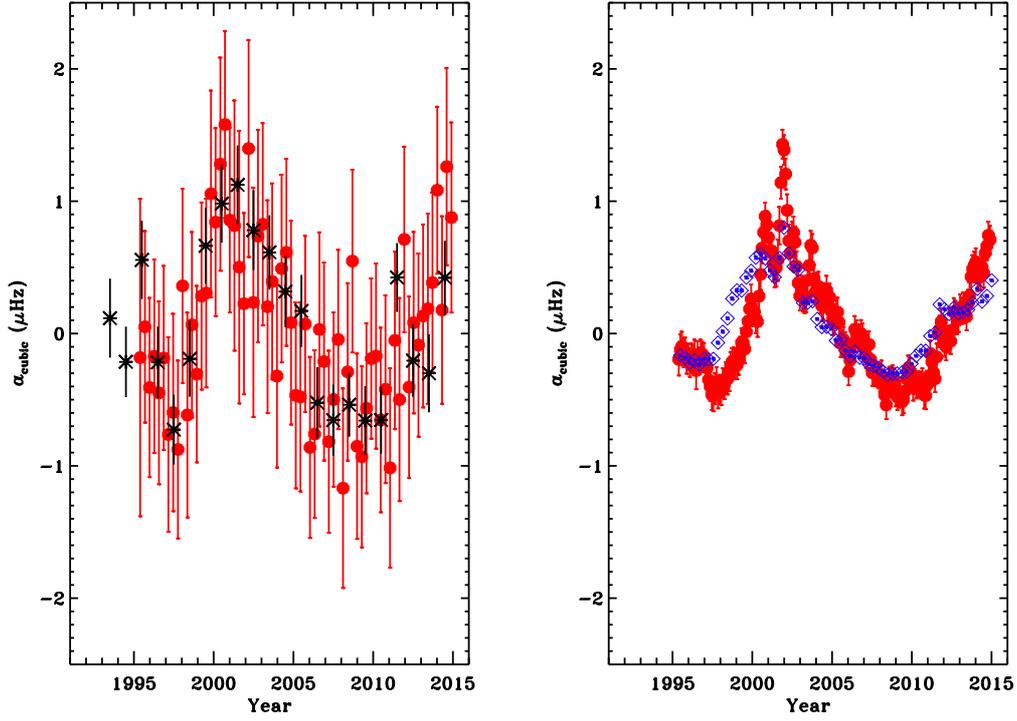}
\caption{Left: Variation of the cubic terms from a two-term fit to the
$l=1,2$ modes only, for BiSON (stars) and GONG sectoral modes only (filled circles). Right: Cubic terms from GONG full multiplets (open diamonds) and sectoral modes only (filled circles) up to $l=150$.}
\label{fig:temp3}
\end{figure*}

We note that the scale of the variation appears to be larger for the low-degree BiSON modes than for the medium-degree modes.
The slope of a linear regression between the two-term $a_{\mathrm{cubic}}$ and the
RF index is 
($13 \pm 1$) nHz per radio flux unit (RFU) for BISON and $(7.59\pm 0.02)$ nHz per RFU for GONG, while for a regression with the magnetic index the corresponding slopes
are $(158\pm 12)$ and $(90.2\pm 0.2)$ nHz/G respectively.  
This does appear to indicate a systematic difference between the two data sets. At first glance this is something of a puzzle. It can be explained by the fact that the BiSON measurements for $l>0$ are heavily weighted to the sectoral ($|m|=l$) spherical harmonic components, while the resolved-Sun measurements sample the whole range of azimuthal order. As the sectoral modes are concentrated towards the equator, their sampling of the activity distribution is not a straightforward latitudinal average and therefore a global activity proxy is not fully appropriate. This was investigated by  \citet{2004MNRAS.352.1102C}. If we restrict the
analysis to the $l=1$ and $l=2$ modes (Fig.~\ref{fig:temp3}) using only the sectoral mode frequencies for GONG, 
we obtain regression slopes of $(12.7\pm 0.2)$ and $(12.2\pm 0.2)$ nHz/RFU for BiSON and GONG respectively, which are much closer to being consistent. 
The remaining discrepancy may be accounted for by the less than perfect nature of the linear relationship between the global activity proxy and the sectoral mode response. In Fig.~\ref{fig:temp3}(b) we also show the variation of the $a_{\mathrm{cubic}}$ coefficient for from sectoral-only GONG frequencies up to $l=150$, plotted on the same axes as those derived from GONG 
central-mode frequencies. This clearly shows that the sectoral modes have a different scale of variation (and a different temporal evolution) from the global averages.
The effect of the latitudinal distribution of magnetic activity on the frequency variation is addressed in more detail in Section~\ref{sec:coeff}.

\subsubsection{Degree dependence for medium-degree modes}

\begin{figure*}
\epsfxsize=0.8\linewidth\epsfbox{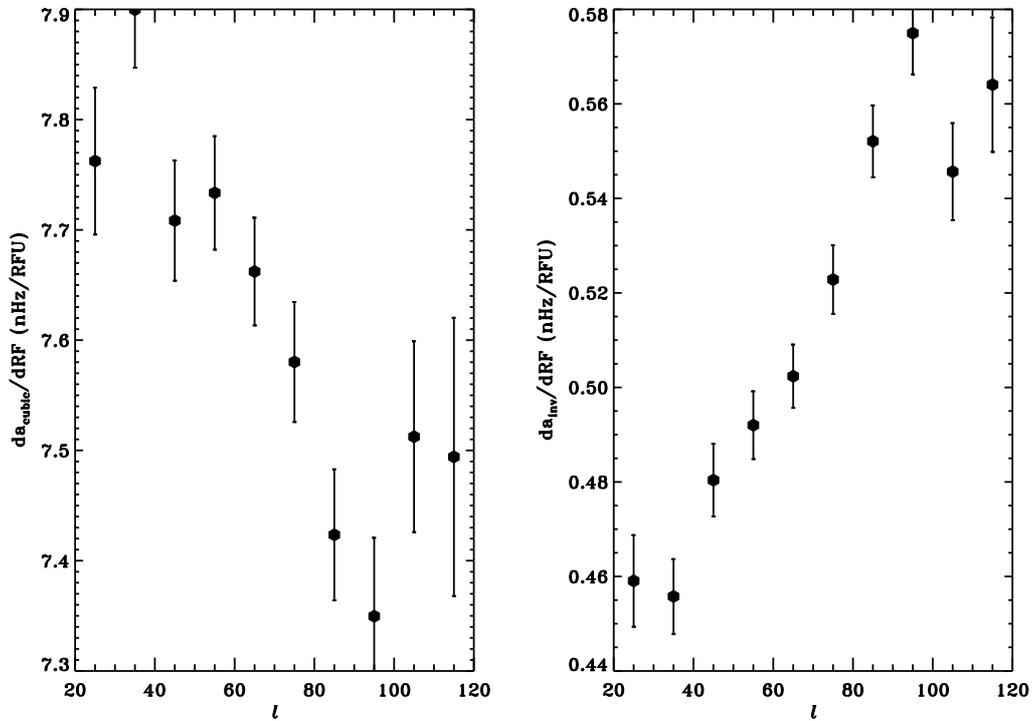}
\caption{Regression slope between RF index and cubic (left) and 
inverse (right) coefficients for  medium-degree 
GONG 
data 
as a function of degree $l$ in 10-$l$ bands.}
\label{fig:temp1}
\end{figure*}

\begin{figure}
\epsfxsize=0.8\linewidth\epsfbox{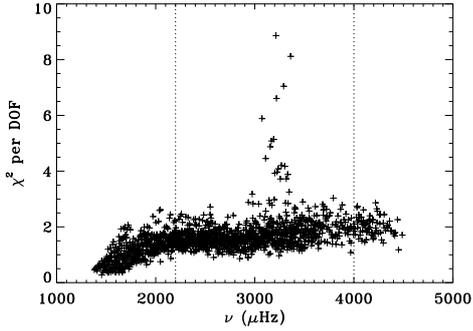}
\caption{$\chi^2$ per degree of freedom for each medium-degree
mode based on the two-term fit to GONG frequency-shift data between $2200$ and $4000\,\mu{{\mathrm{Hz}}}$.
The vertical dotted lines
indicate the limits of the fitting range.}
\label{fig:temp4}
\end{figure}

If we divide the medium-degree values into subsets by $l$, we see that there is a slight $l$-dependence of 
the regression slopes for the cubic and inverse coefficients (Fig.~\ref{fig:temp1}). This is not too surprising, as the simple parametrization was originally designed for low-degree modes only. However, the variation is small enough that we can describe all of the modes by a single pair of parameters to a reasonable approximation. To illustrate this, we show in Fig.~\ref{fig:temp4} the $\chi^2$ per degree of freedom for each mode in the GONG 
sets, where the model is simply the two-term fit to the frequency shifts between $2200$ and $4000\,\mu{{\mathrm{Hz}}}$. Except for a few outliers 
the level of discrepancy between the fit and the individual mode variation
shows little change across a frequency range extending well beyond that from which the fit was derived.

\subsubsection{Correlation between cubic and inverse terms}

We also note that for GONG the cubic and inverse terms are strongly correlated, with a correlation coefficient
0.98 for the 68 non-overlapping GONG samples. 
This does suggest that a different model  or scaling scheme might give an even better fit to the data. This ratio between the coefficients depends on the scaling scheme used; if instead of $E_{nl}$ we use $E_{nl}/E_{n0}$, that is, the mode inertia normalized such that there is no frequency dependence for $l=0$, the fits are only slightly worse but the correlation between cubic and inverse terms becomes an anticorrelation.  
We can take the analysis a step further by fixing the ratio between the two terms; the appropriate value can be found by singular value decomposition applied to a matrix whose columns are the values of $a_{\mathrm{inv}}$ and $a_{\mathrm{cubic}}$. We can then fit an expression 
\begin{equation}
F(\nu)=a_{{\mathrm{fixed}}}[({{\nu}\over{\nu_{\mathrm{ac}}}})^3+c({{\nu}\over{\nu_{\mathrm{ac}}}})^{-1}]
\end{equation}
where $c$ is the constant determined from the singular value decomposition.
The calculation for the least-squares fit in this case is given in Appendix~\ref{sec:app2}. In Figure~\ref{fig:decomp} we show the temporal variation of the variable term $a_{\mathrm{fixed}}$ and the chi-square ratio between the fits with the fixed ratio and those with the two free terms, using a value of 0.0653 for the constant $c$. As for Figure~\ref{fig:fig3}, we also show the best linear fits to the RF and global magnetic index. The correlation coefficients between $a_{\mathrm{fixed}}$ and the activity indices are 0.994 and 0.980 for RF and the magnetic index, respectively.

 Finally, returning to our earlier statement that the relationship between frequency shifts and activity index is not perfectly linear, we show in Figure~\ref{fig:temp3x} the variation of $a_{\mathrm{fixed}}$ with the RF index for BiSON and GONG. A slight curvature and hysteresis are clearly visible; in the BiSON case this is probably dominated by the sectoral mode effect discussed above, but the curvature is still visible in the central-frequency GONG data.

\begin{figure}
\epsfxsize=0.8\linewidth\epsfbox{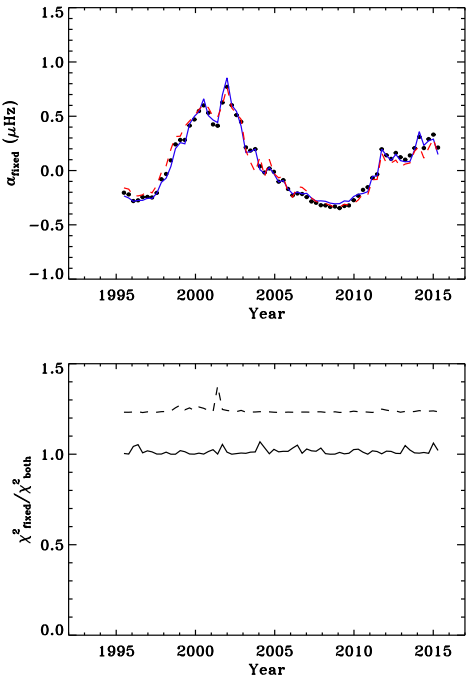}
\caption{Variation of the parameter $a_{\mathrm{fixed}}$ for fits to GONG frequencies (top), with linear fits to the RF index and magnetic index shown as blue and dashed curves respectively, and the ratio (bottom) of $\chi^2$ values ($\chi^2_{\mathrm{fixed}}$) to those for the two-term fit ($\chi^2_{\mathrm{both}}$), with the 0.1 per cent significance level shown as a dashed curve.}
\label{fig:decomp}
\end{figure}

\begin{figure*}
\epsfxsize=0.4\linewidth\epsfbox{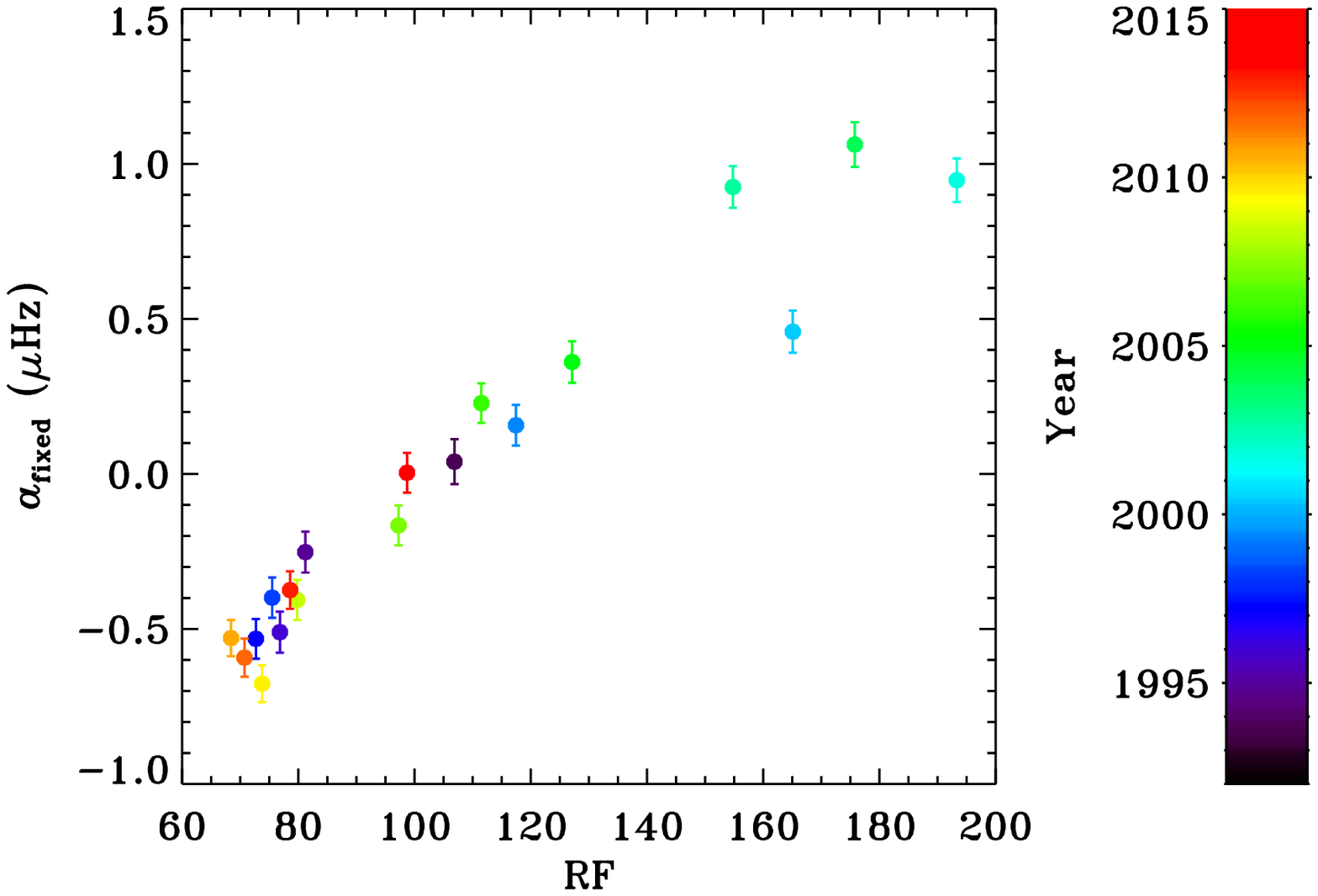}
\epsfxsize=0.4\linewidth\epsfbox{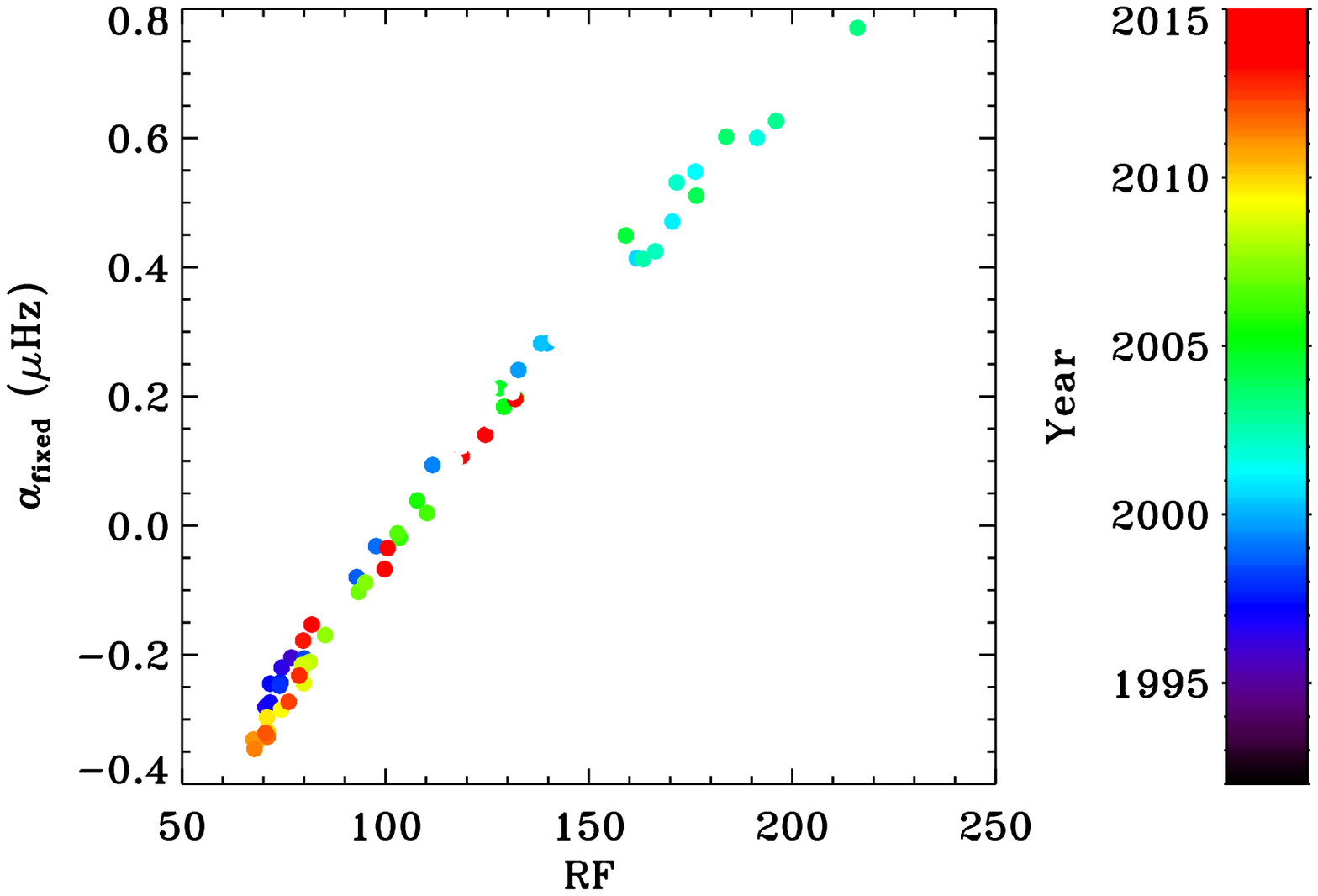}
\caption{Variation of the parameter $a_{\mathrm{fixed}}$ for fits to BiSON (left) and GONG (right) frequencies, colour-coded by year.}
\label{fig:temp3x}
\end{figure*}

\begin{figure*}
\epsfxsize=0.8\linewidth\epsfbox{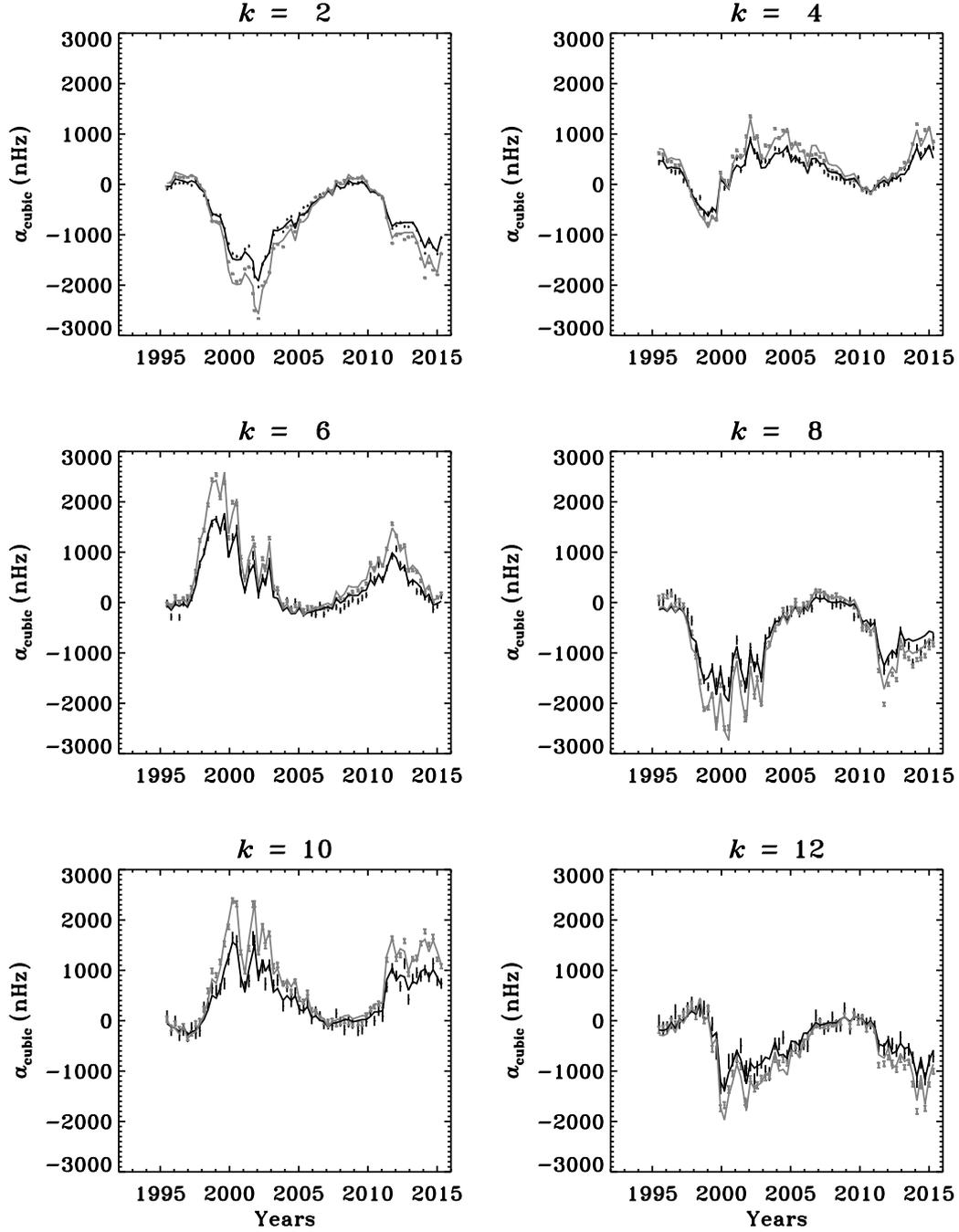}
\caption{Cubic terms from two-term (black) and cubic-only (grey) fits to 
 even-order polynomial expansion coefficients $b_{lk}$ (scaled as described in Equations (3) and (4)) from GONG
data. The solid curves represent the best fit to the corresponding 
magnetic Legendre components $B_{k}$.}
\label{fig:fig4}
\end{figure*}

\begin{figure*}
\epsfxsize=0.8\linewidth\epsfbox{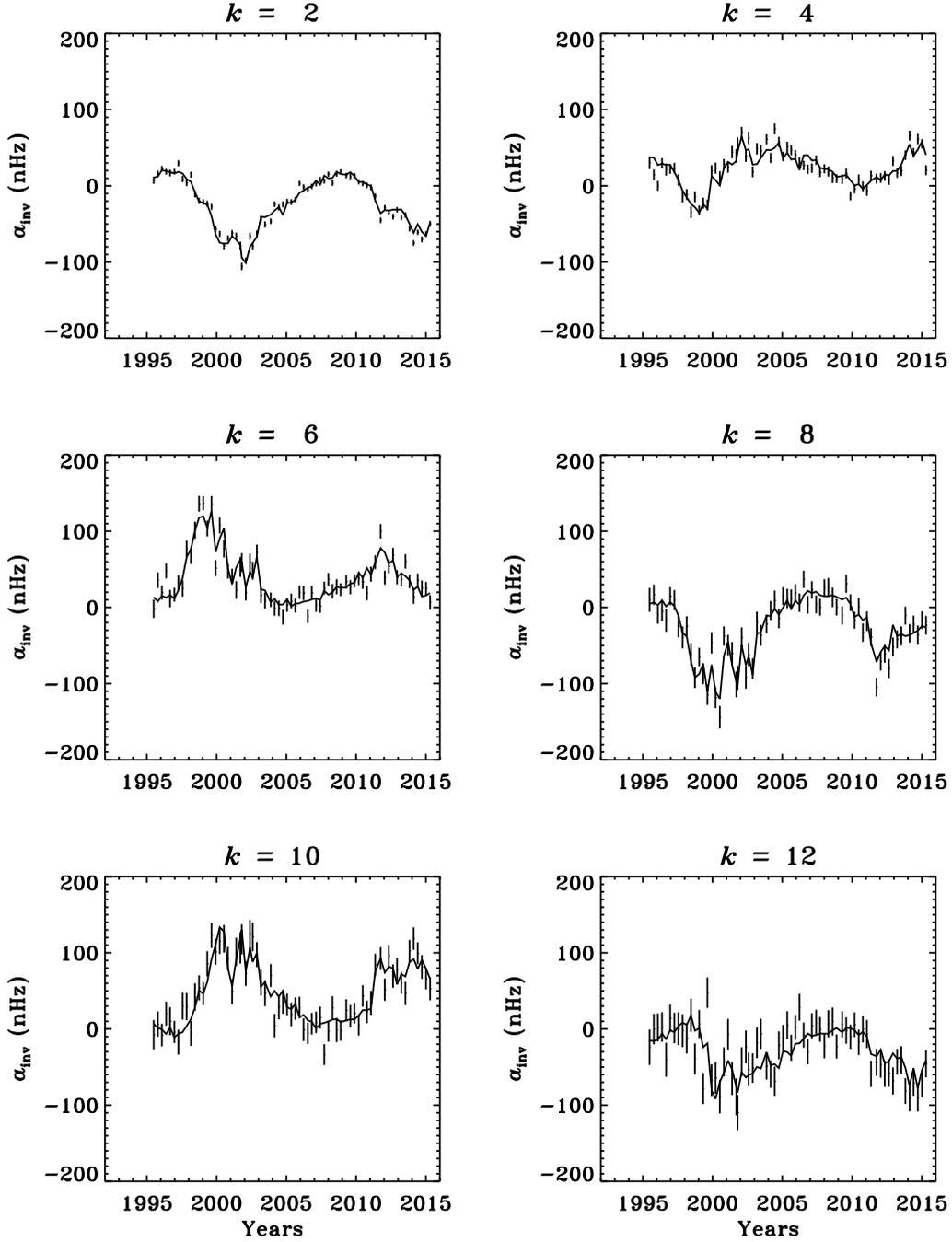}
\caption{Inverse terms from two-term fits to scaled even-order polynomial
expansion coefficients $b_{lk}$ from GONG data.  The solid curves represent the best fit to the corresponding 
magnetic Legendre components $B_{k}$}
\label{fig:fig5}
\end{figure*}

\begin{figure*}
\epsfxsize=0.8\linewidth\epsfbox{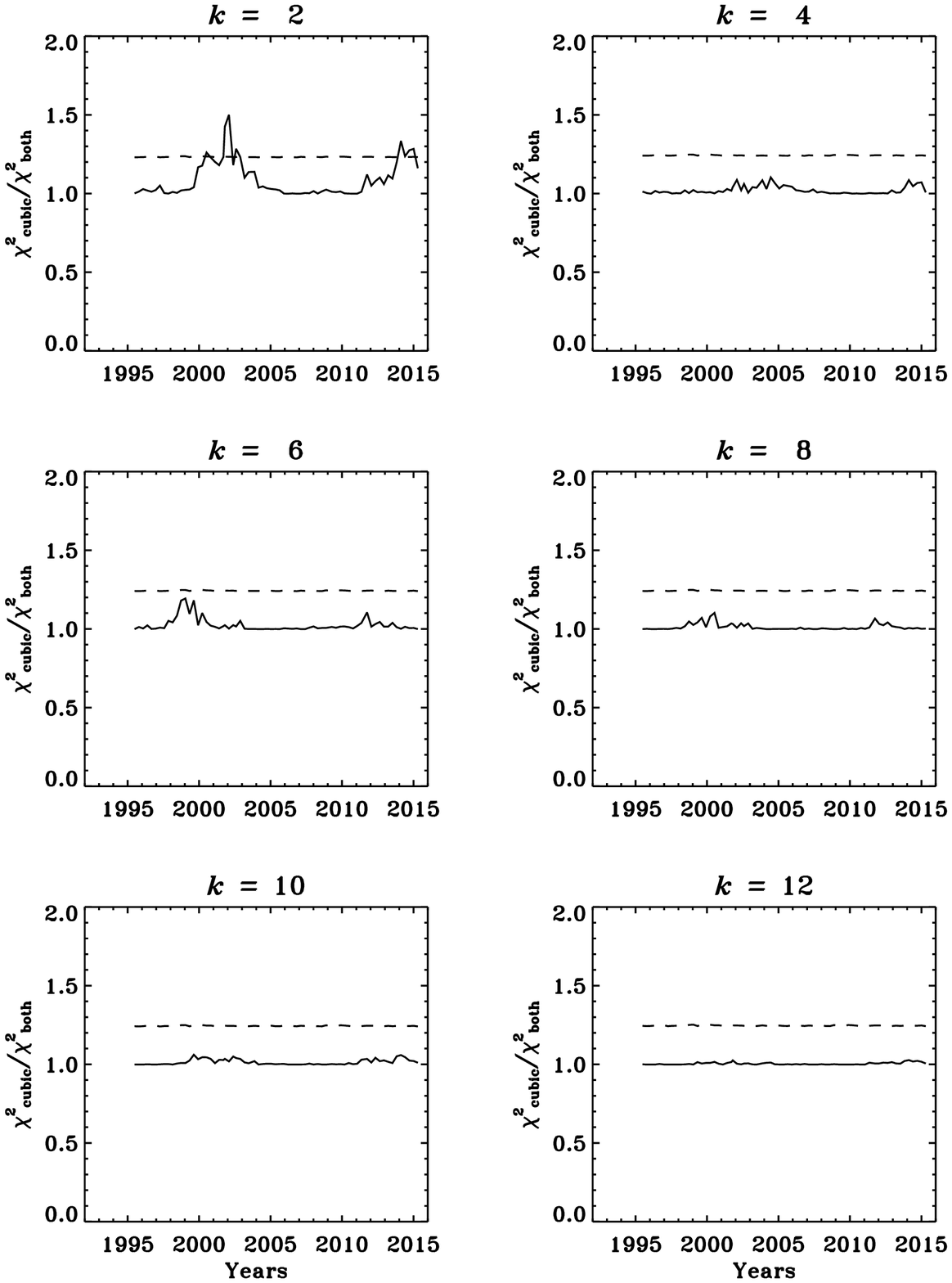}
\caption{Ratio of two-term to cubic-only $\chi^2$ values for fits to 
scaled even-order polynomial expansion coefficients from GONG data.
The dashed curve represents the threshold where the improvement due to the inverse term is significant at the 0.1 per cent level.}
\label{fig:fig6}
\end{figure*}

\subsection{Medium-degree splitting coefficients}
\label{sec:coeff}
In medium-degree data we express the variation of frequency with azimuthal order $m$ for modes of the same radial order and spherical harmonic degree as a polynomial expansion with coefficients $a_k$ for terms of order $k$.  The odd-order coefficients describe the
rotation and the even-order coefficients the `asphericity' -- the deviation from spherical symmetry. Like the frequency shifts, the even-order coefficients show solar-cycle variation due to changes in the near-surface structure, reflecting the changing latitudinal distribution of activity over the cycle. It has been shown that these even-order coefficients are strongly correlated with the Legendre polynomial coefficients describing the latitudinal 
distribution of surface magnetic activity \citep{1999ApJ...524.1084H,2001MNRAS.327.1029A,2002ApJ...580.1172H}. \citet{2001MNRAS.327.1029A}, using coefficients fom GONG and MDI data, showed that if the latitudinal distribution of unsigned surface magnetic field strength (averaged over a Carrington rotation) is expressed as a sum of Legendre polynomials $P_k(\cos\theta)$, where $\theta$ is latitude and $k$ is the polynomial order, with time-varying coefficients $B_k(t)$ such that 
$B(t,\cos\theta)=\sum_k{B_k(t)P_k(\cos\theta)}$, not only are the $B_k$ coefficients linearly related to the corresponding $b_{lk}$ but the scaling factor is more or less independent of $k$. \citet{2001MNRAS.327.1029A} also attempted to carry out two-dimensional structure inversions using the coefficients and found that the variation appeared in the surface term rather than in the depth-dependent structure.

 Following the analysis of \citet{2001MNRAS.327.1029A} we use the 
coefficients as defined by \citet{1991ApJ...369..557R}, where the basis functions are related to the Clebsch-Gordan coefficients $C_{k0lm}^{lm}$ by
\begin{equation}
{\cal P}_k^{(l)}(m) = {l\sqrt{(2l-k)!(2l+k+1)!}\over
(2l)!\sqrt{2l+1}}C_{k0lm}^{lm},
\label{eq:eq2}
\end{equation}
 to form the scaled coefficients $b_{nlk}\equiv la_{nlk}/Q_{lk}$, where $k$ is the order of the expansion coefficient and
$Q_{lk}$ is given by

\begin{equation}
\int_0^{2\pi}\;\dd\phi\;\int_0^\pi \sin\theta\;\dd\theta\;Y_\ellq^m(Y_\ellq^m)^*
P_{2k}(\cos\theta)={1\over\ellq}Q_{\ellq k}{\cal P}_{2k}^{(\ellq)}(m).
\label{eq:eq9}
\end{equation}

While \citet{2001MNRAS.327.1029A} used a weighted mean over the coefficients common to all of the data sets, we will instead describe each dataset by the result of the two-term or cubic-only fit to all of the available coefficients within the frequency range $2200 \leq \nu \leq 4000 \mu{\mathrm Hz}$. We restrict the analysis to coefficients that appear in 80 per cent of the data sets.

Figure~\ref{fig:fig4} shows the $a_{\mathrm{cubic}}$ terms for the cubic-only and 
two-term fits for the even-order coefficients up to $k=12$ from GONG data, together with linear fits to the appropriate coefficients of the magnetic field strength expansion. $k=0$ is not shown as it just corresponds to the central frequency and global magnetic index discussed in Section~\ref{sec:cent}. Figure~\ref{fig:fig5} 
shows the inverse terms from the two-term fits and Figure~\ref{fig:fig6}
shows the $\chi^2$ ratios and thresholds.
Except for $b_2$ during the solar maxima of Cycles 23 and 24, the inverse term is not making a statistically significant contribution to the overall fit; again, the inverse term matters when the signal-to-noise of the coefficients is high, and so it is required to describe the variation overall. As with the central frequencies, the cubic and inverse coefficients show a strong positive correlation. The fixed-ratio fits are not shown, but they would lie almost exactly over the $a_{\mathrm{cubic}}$ values in Figure~\ref{fig:fig4}.

\begin{figure*}
\epsfxsize=0.6\linewidth\epsfbox{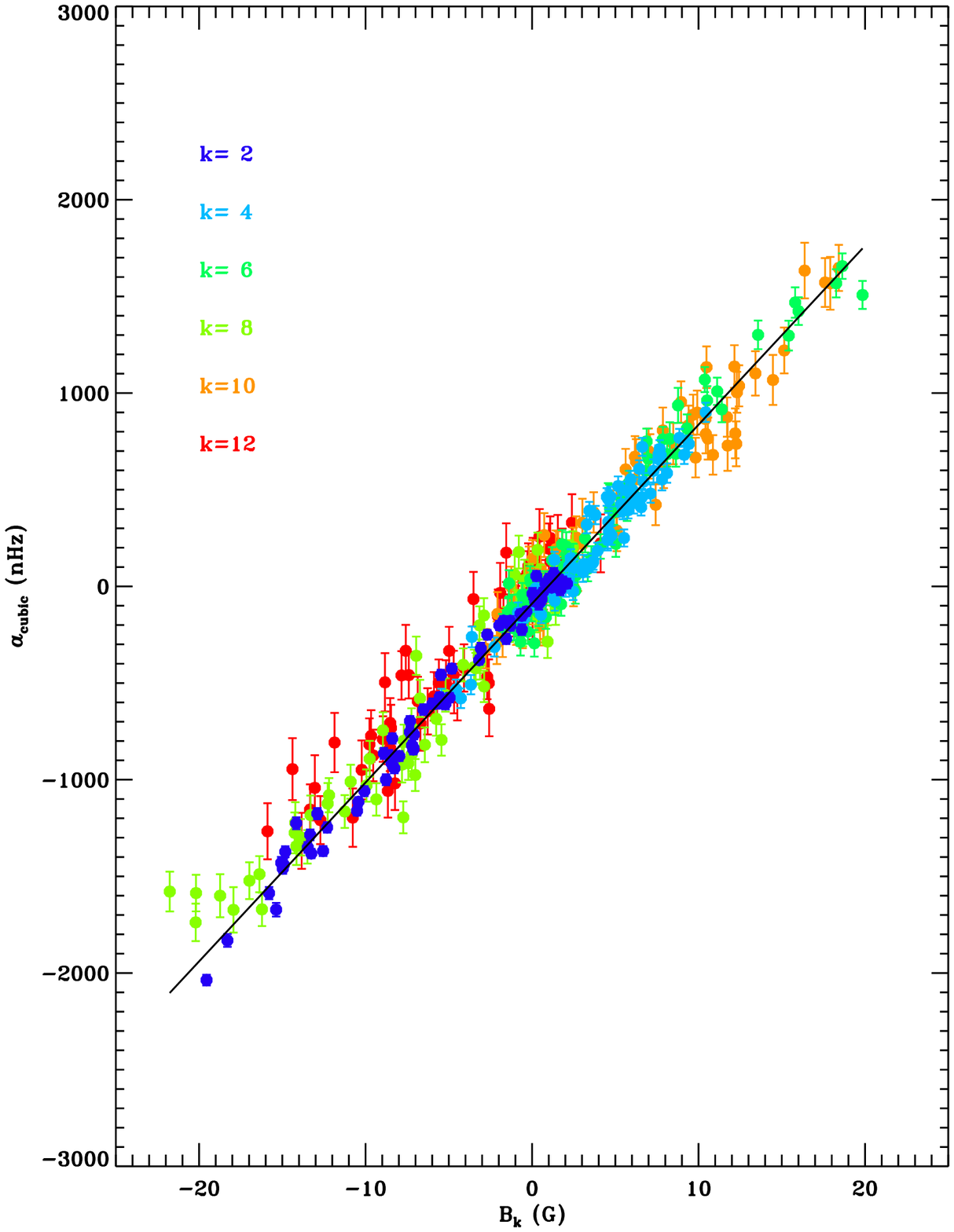}
\caption{The $a_{\mathrm{cubic}}$ terms {\bf from two-term fits} for the GONG $b_{lk}$ scaled  splitting coefficients for even polynomial order $k$ up to $k=12$, plotted against the corresponding magnetic coefficients $B_k$, colour-coded by $k$. The solid line represents a linear least-squares fit to the values for all orders.}
\label{fig:fig7}
\end{figure*}

If we overplot all of the $a_{\mathrm{cubic}}$ values from the two-term fit on the same axes (Fig.~\ref{fig:fig7}) we see that they all lie close to the same approximately straight line. This is consistent with the results of \citet{2001MNRAS.327.1029A} and the earlier findings of \citet{1993ApJ...402L..77W}. We can therefore capture nearly all of the activity-related variation of not only the central frequencies but also the
even-order splitting coefficients using just one or two parameters (the
coefficients describing the dependence of $a_{\mathrm{cubic}}$ and $a_{\mathrm{inv}}$ on the $B_k$ components), or even a single coefficient for $a_{\mathrm{fixed}}$ and a constant $c$, together with the $E_{nl}$ values and the magnetic field distribution.

\section{DISCUSSION AND CONCLUSIONS}
We have attempted to parametrize the the low- and medium-degree frequency shifts in solar p mode frequencies during the solar cycle using mode-inertia scaling together with cubic and inverse frequency terms, as suggested by \citet{2014A&A...568A.123B} for the `surface term' correction.  We find that the cubic term is the more important of the two and could be used alone for the low-degree solar data.


This parametrization suggests a way to quantify possible activity-cycle variations in the mode frequencies of solar-like stars detected by asteroseismology, as it allows us to combine the variation from a large number of modes into a small number of parameters -- and hence reduce the uncertainty in the amplitude of the frequency change -- without knowing the underlying activity index. This would require that the star in question be well enough modelled to provide a suitable mode-inertia table, but this should not present a difficulty, 
as the variation of the mode inertia is small across an ensemble of models that fit the seismic data.

Another useful implication for asteroseismology is that, as both the surface term and the solar-cycle variation can be parametrized using the same expression, the effects of any activity-related varation can be removed in the same correction and should not influence the stellar parameters estimated from modelling.

 While the description is not perfect, the parametrization does capture most of the variation even at medium degree and including a latitudinally-varying activity distribution. The inverse and cubic terms are both required to give a good fit to the medium-degree data when the signal-to-noise in the frequency shifts is high, but when the signal-to-noise is lower the inverse term is not statistically significant even though it may be contributing as much as half of the (small) total shift at low frequencies. The strong correlation between the cubic and inverse terms in the medium-degree data suggests that
we could even use a single variable and a fixed ratio between cubic and inverse terms; however, the value of this ratio is most likely model- or star-dependent and cannot be treated as a universal constant. This is not the only possible way to summarize the variations. For example, \citet{2008ApJ...686.1349B} used a principal-component analysis to remove the surface contribution from the frequency varation in order to search for small changes deep in the convection zone, and they found that a single principal component was sufficient to describe the surface-term variation. However, they did not give a functional form to that component. The component is not linear in frequency, and our work suggests that it would need both the cubic and inverse terms to fit it. In this work, by contrast, we focus on describing the surface term in a physically motivated way.

In future work, we hope to use the parametrization for a more detailed study of differences between the response of low-degree solar frequencies to activity in different solar cycles, and also to apply it to variations in the frequencies of modes from solar-like asteroseismic targets.

\section*{Acknowledgments}
We would like to thank all those who are, or have been, associated with BiSON. BiSON is funded by the Science and Technology Facilities Council (STFC).
This work utilizes data obtained by the Global Oscillation Network Group (GONG) Program, managed by the National Solar Observatory, which is operated by AURA, Inc. under a cooperative agreement with the National Science Foundation. The data were acquired by instruments operated by the Big Bear Solar Observatory, High Altitude Observatory, Learmonth Solar Observatory, Udaipur Solar Observatory, Instituto de Astrof{\'i}sica de Canarias, and Cerro Tololo Interamerican Observatory. NSO/Kitt Peak data used here were produced cooperatively by NSF/NOAO, NASA/GSFC, and NOAA/SEL; SOLIS data are produced cooperatively by NSF/NSO and NASA/LWS.

 R.H. also acknowledges computing support from the National Solar Observatory. W.H.B. acknowledges research funding by the Deutsche Forschungsgemeinschaft (DFG) under grant SFB 963/1 ``Astrophysical flow instabilities and turbulence'' (Project A18). S.B. acknowledges partial support from NSF grant AST-1514676.

\bibliography{ms}

\appendix

\section{TWO-TERM AND ONE-TERM LEAST-SQUARES FITS}
\label{sec:app}

\subsection{Two-term fit calculation}

The two-term expression we use is linear in the coefficients, but because 
there is no constant term we need a custom least-squares fit.

The two-term model is 
\begin{equation}
y={ax^p +bx^q},
\end{equation}
where $p$ and $q$ are chosen constant powers -- $p=3$ and $q=-1$ in the default model.

Hence, for $N$ data points with dependent variable $y_i$, independent variable
$x_i$, and uncertainty $\sigma_i$, we have
\begin{equation}
{\chi^2}={\sum_{i=1}^N{({y_i-(ax_i^p+bx_i^q))^2}\over{\sigma_i^2}}}.
\label{eq:a2}
\end{equation}
For a solution we need 
\begin{equation}
{\partial{\chi^2}\over\partial{a}}={\partial{\chi^2}\over\partial{b}}=0. 
\label{eq:a3}
\end{equation}

Expanding Equation~(\ref{eq:a2}) and taking the partial differentials we 
obtain 
\begin{equation}
{\partial{\chi^2}\over\partial{a}}={-2{\sum_{i=1}^N{{y_ix_i^p-ax_i^{2p}-bx_i^{p+q}}\over{\sigma_i^2}}}},
\end{equation}
and
\begin{equation}
{\partial{\chi^2}\over\partial{b}}={-2{\sum_{i=1}^N{{y_ix_i^q-bx_i^{2q}-ax_i^{p+q}}\over{\sigma_i^2}}}},
\end{equation}

For clarity, we substitute
\begin{eqnarray}
S_1 & ={\sum_{i=1}^N{{y_ix_i^p}\over{\sigma_i^2}}},\label{eq:subs1} \\ 
S_2 & ={\sum_{i=1}^N{{x_i^{2p}}\over{\sigma_i^2}}}, \\
S_3 & ={\sum_{i=1}^N{{x_i^{p+q}}\over{\sigma_i^2}}}, \\
S_4 & ={\sum_{i=1}^N{{y_ix_i^q}\over{\sigma_i^2}}}, \\
S_6 & ={\sum_{i=1}^N{{x_i^{2q}}\over{\sigma_i^2}}}. \label{eq:subs2}
\end{eqnarray}

We can now write Equations~(\ref{eq:a3}) in the form
\begin{equation}
aS_2+bS_3=S_1, aS_3+bS_6=S_4
\end{equation}
and solve for $a$ and $b$ to give 
\begin{eqnarray}
a & ={{S_3S_4-S_1S_6}\over{D}}, \\
b & ={{S_1S_3-S_4S_2}\over{D}},
\end{eqnarray}

where
\begin{equation}
D=S_3^2-S_2S_6.
\end{equation}

The formal uncertainties $\sigma_a, \sigma_b$ can then simply be derived by using
\begin{equation}
\sigma^2_a = \sum_{i=1}^N{{\sigma_i^2}({\partial{a}\over\partial{y_i}})^2},
\sigma^2_b = \sum_{i=1}^N{{\sigma_i^2}({\partial{b}\over\partial{y_i}})^2}.
\end{equation}

with 
\begin{equation}
{\partial{a}\over\partial{y_i}}={{{1\over{D}}({S_3{x_i^q\over{\sigma_i^2}}}-{S_6{x_i^p\over{\sigma_i^2}}}})}
\end{equation}
and
\begin{equation}
{\partial{b}\over\partial{y_i}}={{{1\over{D}}({S_3{x_i^p\over{\sigma_i^2}}}-{S_2{x^q\over{\sigma_i^2}}}})}.
\end{equation}

\subsection{Single-term fit}
In the single-term case the calculation is simpler. The model is just
$y=a^p$, so 
\begin{equation}
\chi^2={\sum_{i=1}^N{({y_i-ax_i^p)^2}\over{\sigma_i^2}}},
\end{equation}
which yields 
\begin{equation}
a={{{\sum_{i=1}^N{y_ix^p\over{\sigma_i^2}}}}\over{\sum_{i=1}^N{{x^{2p}}\over{\sigma_i^2}}}}
\end{equation}
and 
\begin{equation}
{\partial{a}\over\partial{y_i}}={{{x_i^p}\over{\sigma_i^2}}\over{\sum_{i=1}^N{{x^{2p}\over{\sigma_i^2}}}}}.
\end{equation}

\section{SINGLE-PARAMETER TWO-TERM FIT}
\label{sec:app2}
If we fix the ratio between the two power-law terms, the
expression becomes
\begin{equation}
y=a(x^p+cx^q), 
\end{equation}
where $c$ is a constant term obtained from singular value decomposition of the results of the two-term fit.

In this case, 
\begin{equation}
\chi^2=\sum_{i=1}^N{{(y_i-(ax_i^p+acx_i^q))^2}\over{\sigma_i^2}.}
\end{equation}

As before, we expand and take the partial differential to solve $\partial\chi^2/\partial a=0$ for $a$;
\begin{equation}
{\partial\chi^2/\partial a}=2{\sum_{i=1}^N}{{a(x^{2p}+2cx^{p+q}+c^2x^{2*q})-y(x^p+cx^q)}\over\sigma_i^2} \equiv 0.
\end{equation}

Using the same substitutions as in Equations~\ref{eq:subs1}\,--\,\ref{eq:subs2},
this reduces to 
 \begin{equation}
a={{S_1+cS_4}\over D},
\end{equation}
where 
\begin{equation}
D=S_2+2cS_3+c^2S_6.
\end{equation}

The derivative to obtain the error on $a$ is
\begin{equation}
{{\partial a}\over{\partial y_i}}={1\over D}{{x_i^p+cx_i^q}\over{\sigma_i^2}}.
\end{equation}

\label{lastpage}

\end{document}